\documentstyle[12pt,epsfig]{article}

\newskip\humongous \humongous=0pt plus 1000pt minus 1000pt

\newif\ifdtup

\def\oldreffmt#1{\rlap{[#1]} \hbox to 2\parindent{}}

\def\figfmt#1{\rlap{Figure {#1}} \hbox to 1in{}}

\def\beq{\begin{equation}}
\def\eeq{\end{equation}}
\def\bea{\begin{eqnarray}}
\def\eea{\end{eqnarray}}
\def\half{\frac{1}{2}}

\def\bq{\begin{quote}}
\def\eq{\end{quote}}

\def\half{\frac{1}{2}}     

\relax

\newskip\humongous \humongous=0pt plus 1000pt minus 1000pt

\newif\ifdtup

\def\oldreffmt#1{\rlap{[#1]} \hbox to 2\parindent{}}

\def\figfmt#1{\rlap{Figure {#1}} \hbox to 1in{}}

\def\beq{\begin{equation}}
\def\eeq{\end{equation}}
\def\bea{\begin{eqnarray}}
\def\eea{\end{eqnarray}}
\def\half{\frac{1}{2}}

\def\bq{\begin{quote}}
\def\eq{\end{quote}}

\def\half{\frac{1}{2}}     

\relax
\catcode`\@=11

\def\@normalsize{\@setsize\normalsize{15pt}\xiipt\@xiipt
\abovedisplayskip 14pt plus3pt minus3pt%
\belowdisplayskip \abovedisplayskip
\abovedisplayshortskip  \z@ plus3pt%
\belowdisplayshortskip  7pt plus3.5pt minus0pt}

\def\small{\@setsize\small{13.6pt}\xipt\@xipt
\abovedisplayskip 16pt plus3pt minus3pt%
\belowdisplayskip \abovedisplayskip
\abovedisplayshortskip  \z@ plus3pt%
\belowdisplayshortskip  7pt plus3.5pt minus0pt
\def\@listi{\parsep 4.5pt plus 2pt minus 1pt
            \itemsep \parsep
            \topsep 9pt plus 3pt minus 3pt}}

\def\underline#1{\relax\ifmmode\@@underline#1\else
	$\@@underline{\hbox{#1}}$\relax\fi}
\@twosidetrue

\catcode`@=12
     
\begin{document}
\begin{flushright}
\vspace*{0.5in}
\vspace{1in}
Fermilab-FN-687 \\
CDF/ANAL/EXOTIC/PUBLIC/5064 \\
\today \\
\end{flushright}
\begin{center}
\vspace{.5in}
{\LARGE Cross Section for Topcolor $Z'_t$ decaying to $t\bar{t}$}\\
\vspace{.5in}
{\large Version 2.6}\\
\vspace{.5in}
{\large Robert M. Harris, Christopher T. Hill}\\
{\large and Stephen J. Parke}\\
{\large \em Fermilab}\\
\vspace{.2in}
\end{center}
\vspace{.5in}
\begin{abstract}
We present a calculation of the cross section for the process
$p\bar{p} \rightarrow Z'_t \rightarrow t\bar{t}$, the 
production of Topcolor $Z'_t$ with subsequent decay 
to $t\bar{t}$ in $p\bar{p}$ collisions at $\sqrt{s}=1.8$ TeV.
Variations of the cross section with varying assumptions about the model, the 
resonance width, the parton distributions and the renormalization scale 
are presented.
\end{abstract}
\clearpage

\section{Topcolor}

The large mass of the top quark suggests that the third generation may
play a special role in the dynamics of
electroweak symmetry breaking. 
Most models in which this occurs are
based upon  topcolor \cite{ref_topc1,ref_topc2}, which can
generate a large top quark mass 
through the formation of a dynamical $t\bar{t}$ condensate,
generated 
by a new strong gauge force coupling 
preferentially to the third generation. 

In a typical topcolor scheme the QCD gauge group,
 $SU(3)_C$, is imbedded into
a larger structure, e.g.,  $SU(3)_1\times SU(3)_2$
with couplings $h_1$ and $h_2$ respectively.
$SU(3)_2$ ($SU(3)_1$) couples to
the third (second and first) generation,
and $h_2 >> h_1$.  
The breaking  $SU(3)_1\times SU(3)_2
\rightarrow SU(3)_C$  produces a
massive color octet of bosons, known as ``topgluons'', which couple
mainly to $b\bar{b}$ and $t\bar{t}$.  By itself, this scheme would
produce a degenerate top quark and bottom quark.
Moreover, if the condensates were required to
account for all of EWSB, and without excessive fine-tuning,
then the resulting fermion masses  would
 be quite large, $\sim 600$ GeV.

To
get the correct scale of the top quark mass one typically
considers topcolor in tandem with something else, either
an explicit Higgs boson, SUSY, or most naturally
with additional strong dynamics, as in ``topcolor assisted technicolor''
\cite{ref_topc1}. However, another strategy, which seems very promising,
is to invoke a topquark seesaw \cite{dob}. In the latter case, the
topquark condensate does lead {\em ab initio } to a top mass of $\sim 600$ GeV,
but through mixing with other electroweak singlet, vector-like
fermions the physical top mass is ``seesawed'' down to its physical value.
Again, it is the heaviness of the top quark that makes this latter
scheme natural, and minimizes fine tuning. The top quark seesaw seems to 
emerge naturally in extensions to extra
space-time  dimensions at the TeV scale \cite{dht}.

Clearly, all such models require yet
another component.  
Indeed, a ``tilting'' mechanism is required to enhance the
formation of the $t\bar{t}$ condensate, while blocking the
formation of the $b\bar{b}$ condensate in all such
schemes so that the b-quark is light while top
is heavy.  This tilting mechanism is
constrained by the $\rho$--parameter (or T parameter) because
it clearly must violate custodial $SU(2)$

One way to provide the tilting
mechanism is to introduce a neutral gauge boson, $Z'$, 
with an attractive interaction between $t\bar{t}$
and a repulsive interaction between $b\bar{b}$. In fact,
the $Z$ boson of the Standard Model does precisely this
and could itself provide the tilting, however the SM coupling constant
$g_1$ is so small that one would be fine-tuning to achieve tilting
in the presence of a large $h_2$.  Hence, typically we
introduce a new $Z'$ boson to drive the tilting.  

There are many ways to engineer the tilting with a new $Z'$.
Obviously anomaly cancellation is mandated for all gauge
forces, but this is not a sufficiently powerful constraint to
uniquely specify the couplings.  The simplest approach
is to imbed $U(1) \rightarrow U(1)_1\times U(1)_2$ in complete
analogy to the topcolor imbedding, and each $U(1)_i$ is
just the appropriate weak hypercharge operator, with $i=2$ ($i=1$)
acting on the third (second and first) generation.  This produces
a topcolor $Z'$, the $Z'_t$, which couples strongly to the third
generation and weakly to the first and second, and which, remarkably,
can satisfy all of the constraints of flavor changing
processes \cite{bbhk} (despite the loss of explicit GIM cancellation). 

In the present paper we will consider the physics in production
and decay of the $Z'_t$.  In addition to the standard $Z'_t$ discussed above,
which we call Model I, we will present three additional new 
models of the $Z'_t$ (Model's II, III and IV). We will find that the standard
$Z'_t$ from Model I has the lowest production cross section of the four 
models. Although the standard $Z'_t$ could be found in this decay channel 
at the Fermilab Tevatron Collider beginning in the next run, it is more likely 
to be seen first in the leptonic decay mode at the Tevatron.  
Models II and III are similar to Model I but yield a higher cross section in 
the $t\bar{t}$ decay channel. The $Z'_t$ from Model IV represents a novel 
class of solutions to the tilting problem. It couples strongly only to the 
first and third generation of quarks. This $Z'_t$ from Model IV has no 
significant couplings to leptons. It is therefore leptophobic and topophyllic.

\section{Topcolor $Z'$ Models}

\noindent
{\bf \Large Standard $Z'_t$ Production and Decay}

\noindent
({\bf{Model I}}): 
generation $(3) \supset U(1)_2$ and generations $(1,2)\supset U(1)_1$

\vskip 0.05in

\noindent
We consider
incoherent production, which does not include
$\gamma-Z-Z'_t$ interference terms.  This is
valid in the narrow width
approximation for $Z'_t$. 
We use a convention of spin-summing
and color-summing both initial and final states.
This requires a color-averaged and spin-averaged structure function.
 
The interaction Lagrangian for
the $Z'_t$ first proposed
in \cite{ref_topc1} is {\bf{Model I}}:
\bea
L' & = & 
(\half g_1\cot\theta_H)Z_t'^{\mu}\left(  
\frac{1}{3}(\bar{t}_L\gamma_\mu t_L 
+\bar{b}_L\gamma_\mu b_L)
+\frac{4}{3} \bar{t}_R\gamma_\mu  t_R 
-\frac{2}{3} \bar{b}_R\gamma_\mu  b_R 
\right.\nonumber \\
& & \left.
- \bar{\tau}_L\gamma_\mu  \tau_L - \bar{\nu_{\tau L}}\gamma_\mu \nu_{\tau L}
- 2\bar{\tau}_R\gamma_\mu  \tau_R 
\right)\nonumber \\
& & 
-(\half g_1\tan\theta_H)Z_t'^{\mu}\left(  
\frac{1}{3}(\bar{u}_L\gamma_\mu u_L 
+ \bar{d}_L\gamma_\mu d_L)
+\frac{4}{3} \bar{u}_R\gamma_\mu  u_R 
-\frac{2}{3} \bar{d}_R\gamma_\mu  d_R 
\right.\nonumber \\
& & \left.
- \bar{e}_L\gamma_\mu  e_L - \bar{\nu_{e L}}\gamma_\mu \nu_{e L}
- 2\bar{e}_R\gamma_\mu  e_R 
\right)\nonumber \\
& & + \left( \makebox{ second generation $\equiv$ first generation }\right)
\eea
We compute the total cross-section
$\sigma(q\bar{q} \rightarrow Z'_t \rightarrow t\bar{t})$
keeping the top quark mass dependence
and spin-summing and color-summing on both
initial and final states:
\bea
\sigma (u+\bar{u}\rightarrow t+\bar{t})
& = &
\frac{9 \alpha^2 \pi  }{16 \cos^4\theta_W}
\left(\frac{17}{9} \right)
\left[ \beta(1 +\frac{1}{3}\beta^2)
\left(\frac{17}{9} \right)
+ \frac{8}{9}\beta (1-\beta^2)
\right]\times
\nonumber \\
& & \qquad 
\times \left[ \frac{s}{(s-M_{Z'_t}^2)^2 + s\Gamma^2 } \right]\theta(s-4m_t^2)
\nonumber \\
\eea
and:
\bea
\sigma (d+\bar{d}\rightarrow t+\bar{t})
& = &
\frac{9 \alpha^2 \pi  }{16\cos^4\theta_W}
\left( \frac{5}{9} \right)
\left[ \beta(1 +\frac{1}{3}\beta^2)
\left(\frac{17}{9} \right)
+ \frac{8}{9}\beta (1-\beta^2)
\right] \times
\nonumber \\
& & \qquad 
\times \left[ \frac{s}{(s-M_{Z'_t}^2)^2 + s\Gamma^2 } \right]\theta(s-4m_t^2)
\nonumber \\
\eea
and in general:
\bea
\sigma
& = &
\frac{9 \alpha^2 \pi  }{16\cos^4\theta_W}
\left(\frac{17}{9} \; \makebox{for $u+\bar{u}$;}, 
\; \frac{5}{9} \;\makebox{for $d+\bar{d}$ },
\; \frac{5}{3} \;\makebox{for $e+\bar{e}$ or $\mu+\bar{\mu}$ }
\right)\times
\nonumber \\ 
& & \times
\left[ \beta(1 +\frac{1}{3}\beta^2)
\left(\frac{17}{9} \; \makebox{for $t+\bar{t}$;}\right) 
+\left(\frac{4}{3} \right) \left(\; \frac{5}{9} \;\makebox{for $b+\bar{b}$ };
\; \frac{5}{3} \;\makebox{for $\tau+\bar{\tau}$ };
\; \frac{1}{3} \;\makebox{for $\nu_\tau+\bar{\nu_\tau}$ };
\right)
\right.
\nonumber \\ 
& & \left. \qquad \qquad
+ \frac{8}{9}\beta (1-\beta^2) \; \makebox{(for $t+\bar{t}$)}
\right]\times \nonumber \\
& & \qquad  \times   
\left[ \frac{s}{(s-M_{Z'_t}^2)^2 + s\Gamma^2 } \right]\theta(s-4m_t^2)
\nonumber \\
\label{eq_sigma_I}
\eea

\noindent
We obtain the $Z'_t $  partial decay width to top pairs:
\beq
\Gamma(Z'_t\rightarrow t\overline{t}) = 
\frac{ \alpha\cot^2\theta_H }{8 \cos^2\theta_W } \sqrt{ M_{Z_t'}^2 - 4m_t^2 } 
\left( \frac{17}{9} 
\left[ 1-\frac{m_t^2}{M_{Z'_t}^2} 
\right] 
- \frac{8}{3} 
\left[ \frac{m_t^2}{M_{Z'_t}^2} \right]  \right)
\eeq
The partial width to bottom pairs
and $\tau$ and $\nu_\tau$ (in the limit $m_b\rightarrow 0$):
\beq
\Gamma(Z'_t\rightarrow b\overline{b}) = 
\frac{ \alpha\cot^2\theta_H }{8 \cos^2\theta_W }   
\left( \frac{5}{9}_b + \frac{5}{3}_\tau +  \frac{1}{3}_{\nu\tau}
\right)M_{Z'_t}.
\eeq
The partial width to first [or
second generation] (in the limit $m_b\rightarrow 0$):
\beq
\Gamma(Z'_t\rightarrow q\overline{q}+\ell\overline{\ell}) = 
\frac{ \alpha\tan^2\theta_H }{8 \cos^2\theta_W }   
\left( \frac{17}{9} + \frac{5}{9} + \frac{5}{3} +  \frac{1}{3}
\right)M_{Z'_t}.
\eeq
and hence the total width:
\bea
\Gamma(Z'_t) & = &
\frac{ \alpha M_{Z'_t} \cot^2\theta_H }{8 \cos^2\theta_W }
\left[
 \sqrt{ 1 - \frac{4m_t^2}{M_{Z'_t}^2 }} 
\left( \frac{17}{9} - \frac{41 m_t^2}{9 M_{Z'_t}^2} 
\right) 
+
\left( \frac{5}{9}_b + \frac{5}{3}_\tau +  \frac{1}{3}_{\nu\tau}\right)
+
\right.
\nonumber \\
&  & \left. \qquad \qquad
2\tan^4\theta_H 
\left( \frac{17}{9} + \frac{5}{9} + \frac{5}{3} +  \frac{1}{3}
\right)
\right]
\nonumber \\
& = & \frac{ \alpha M_{Z'_t} \cot^2\theta_H }{8 \cos^2\theta_W }
\left[
 \sqrt{ 1 - \frac{4m_t^2}{M_{Z'_t}^2 }} 
\left( \frac{17}{9} - \frac{41 m_t^2}{9 M_{Z'_t}^2} 
\right) 
+
\left( \frac{23}{9} \right)
+
\tan^4\theta_H 
\left( \frac{80}{9}
\right)
\right]
\label{eq_gamma_I}
\eea

\vskip 0.5in
\noindent
\section{ Non-Standard Topcolor $Z'_t$ Production and Decay}
\vskip 0.05in
\noindent
{\bf \Large (A) Generalized $Z'_t$ production cross-section}
\vskip 0.05in
\noindent
Non-standard models can be constructed in which
the $U(1)_Y\rightarrow U(1)_1\times U(1)_2 $
and the generations are grouped differently: 

\noindent
({\bf{Model II}}): 
generations $(1,3) \supset U(1)_2$ and generation $(2)\supset U(1)_1$

\noindent
({\bf{Model III}}): 
or generations $(1,2,3) \supset U(1)_2$ (analogue of 
Chivukula-Cohen-Simmons \cite{ccs} spectator coloron scheme)

\noindent
as distinct from the usual topcolor in
which generations $(1,2) \supset U(1)_1$ and generation $(3)\supset U(1)_2$.
If $h_2 \gg h_1$, then $\cot\theta_H \gg 1$ 
and this preserves the desirable features
of having a strong $U(1)$ tilting interaction for
the top mass, and now the production of $Z'_t$ from 
first generation fermions is enhanced; we'll neglect
limits on such a new object from radiative corrections
to $Z$ decay, etc.).

We use a convention of spin-summing
and color summing (not averaging) both initial and final states.
This requires a color-averaged and spin-averaged structure function.

\noindent 
The dominant part of the interaction Lagrangian for
{\bf{Model II}} is:
\bea
L'{}_{\bf{II}} & = & 
(\half g_1\cot\theta_H)Z_t'^{\mu}\left(  
\frac{1}{3}\bar{t}_L\gamma_\mu t_L 
+\frac{1}{3}\bar{b}_L\gamma_\mu b_L
+\frac{4}{3} \bar{t}_R\gamma_\mu  t_R 
-\frac{2}{3} \bar{b}_R\gamma_\mu  b_R 
\right. \nonumber \\
& &  
+\frac{1}{3}\bar{u}_L\gamma_\mu u_L 
+\frac{1}{3}\bar{d}_L\gamma_\mu d_L
+\frac{4}{3} \bar{u}_R\gamma_\mu  u_R 
-\frac{2}{3} \bar{d}_R\gamma_\mu  d_R 
\nonumber \\
& &
\left.
- \bar{\tau}_L\gamma_\mu  \tau_L - \bar{\nu_{\tau L}}\gamma_\mu \nu_{\tau L}
- 2\bar{\tau}_R\gamma_\mu  \tau_R 
- \bar{e}_L\gamma_\mu  e_L - \bar{\nu_{e L}}\gamma_\mu \nu_{e L}
- 2\bar{e}_R\gamma_\mu  e_R 
\right)
\eea
\noindent
The dominant part of the interaction Lagrangian for
{\bf{Model III}} is:
\bea
L'{}_{\bf{III}} & = & 
(\half g_1\cot\theta_H)Z_t'^{\mu}\left(  
\frac{1}{3}\bar{t}_L\gamma_\mu t_L 
+\frac{1}{3}\bar{b}_L\gamma_\mu b_L
+\frac{4}{3} \bar{t}_R\gamma_\mu  t_R 
-\frac{2}{3} \bar{b}_R\gamma_\mu  b_R 
\right. \nonumber \\
& & 
\left.  
+\frac{1}{3}\bar{u}_L\gamma_\mu u_L 
+\frac{1}{3}\bar{d}_L\gamma_\mu d_L
+\frac{4}{3} \bar{u}_R\gamma_\mu  u_R 
-\frac{2}{3} \bar{d}_R\gamma_\mu  d_R 
\right. \nonumber \\
& &  
+\frac{1}{3}\bar{c}_L\gamma_\mu u_L 
+\frac{1}{3}\bar{s}_L\gamma_\mu d_L
+\frac{4}{3} \bar{c}_R\gamma_\mu  u_R 
-\frac{2}{3} \bar{s}_R\gamma_\mu  d_R 
\nonumber \\
& & 
- \bar{\tau}_L\gamma_\mu  \tau_L - \bar{\nu_{\tau L}}\gamma_\mu \nu_{\tau L}
- 2\bar{\tau}_R\gamma_\mu  \tau_R 
- \bar{e}_L\gamma_\mu  e_L - \bar{\nu_{e L}}\gamma_\mu \nu_{e L}
- 2\bar{e}_R\gamma_\mu  e_R 
\nonumber \\
& & \left.
- \bar{\mu}_L\gamma_\mu  \mu_L - \bar{\nu_{\mu L}}\gamma_\mu \nu_{\mu L}
- 2\bar{\mu}_R\gamma_\mu  \mu_R \right)
\eea
The non-standard $Z'_t$ production cross-section
$\sigma(q\bar{q} \rightarrow Z'_t \rightarrow t\bar{t})$
is kinematically identical to the
standard $Z'_t$ case discussed above. The results are:
\bea
\sigma_{\bf{II}}
& = &
\frac{9 \alpha^2 \pi  }{16\cos^4\theta_W}\cot^4\theta_H
\times \left(\frac{17}{9} \; \makebox{for initial
state $u+\bar{u}$;}, 
\; \frac{5}{9} \;\makebox{for initial $d+\bar{d}$ }\right)
\nonumber \\ 
& & \times
\left[ \beta(1 +\frac{1}{3}\beta^2)
 \times \left(\frac{17}{9} \; \makebox{for final $t+\bar{t}$ or $u+\bar{u}$;}, 
\; \frac{5}{9} \;\makebox{for final  $b+\bar{b}$ or $d+\bar{d}$ }\right)
\right.
\nonumber \\ 
& & + \left. \frac{8}{9}\beta (1-\beta^2) \;  \makebox{(for
final  $t+\bar{t}$)}
\right] 
\left[ \frac{s}{(s-M_{Z'_t}^2)^2 + s\Gamma^2 } \right]\theta(s-4m_t^2)
\nonumber \\
\eea

\bea
\sigma_{\bf{III}}
& = &
\frac{9 \alpha^2 \pi  }{16\cos^4\theta_W}\cot^4\theta_H
\times  \left(\frac{17}{9} \; \makebox{for initial $u+\bar{u}$;}, 
\; \frac{5}{9} \;\makebox{for initial $d+\bar{d}$ }\right)
\nonumber \\ 
& & \times
\left[ \beta(1 +\frac{1}{3}\beta^2)
\times \left(\frac{17}{9} \; \makebox{for final $t+\bar{t}$ or $u+\bar{u}$
or $c+\bar{c}$ ;}, 
\; \frac{5}{9} \;\makebox{for final $b+\bar{b}$ or $d+\bar{d}$ 
or $s+\bar{s}$ }\right)
\right.
\nonumber \\ 
& & + \left. \frac{8}{9}\beta (1-\beta^2) \; \makebox{(for 
final $t+\bar{t}$)}
\right] 
\left[ \frac{s}{(s-M_{Z'_t}^2)^2 + s\Gamma^2 } \right]\theta(s-4m_t^2)
\nonumber \\
\eea
The decay kinematics are the same as
for standard $Z'_t$.
Hence, for Model
{\bf{II}}:
\beq
\Gamma_{\bf{II}}(Z'_t\rightarrow t\overline{t}) = 
\frac{ \alpha\cot^2\theta_H }{8 \cos^2\theta_W } \sqrt{ M_{Z'_t}^2 - 4m_t^2 } 
\left( \frac{17}{9} 
\left[ 1-\frac{m_t^2}{M_{Z'_t}^2} 
\right] 
- \frac{8}{3} 
\left[ \frac{m_t^2}{M_{Z'_t}^2} \right]  \right)
\eeq

\beq
\Gamma_{\bf{II}}(Z'_t\rightarrow u\overline{u}) = 
\frac{ \alpha\cot^2\theta_H }{8 \cos^2\theta_W } M_{Z'_t}
\left( \frac{17}{9} 
\right)
\eeq

\beq
\Gamma_{\bf{II}}(Z'_t\rightarrow b\overline{b} \; or \; d\overline{d}) = 
\frac{ \alpha\cot^2\theta_H }{8 \cos^2\theta_W }   
\left( \frac{5}{9} 
\right)M_{Z'_t}.
\eeq

\beq
\Gamma_{\bf{II}}(Z'_t\rightarrow e\overline{e} \; or \; \tau\overline{\tau}) = 
\frac{ \alpha\cot^2\theta_H }{8 \cos^2\theta_W }   
\left( \frac{5}{3} 
\right)M_{Z'_t}.
\eeq

\beq
\Gamma_{\bf{II}}(Z'_t\rightarrow \nu_e\overline{\nu_e} \; or \; 
\nu_\tau\overline{\nu_\tau}) = 
\frac{ \alpha\cot^2\theta_H }{8 \cos^2\theta_W }   
\left( \frac{1}{3} 
\right)M_{Z'_t}.
\eeq

\beq
\Gamma_{\bf{II}}(Z'_t\rightarrow c\overline{c}) = 
\frac{ \alpha\tan^2\theta_H }{8 \cos^2\theta_W } M_{Z'_t}
\left( \frac{17}{9} 
\right)
\eeq

\beq
\Gamma_{\bf{II}}(Z'_t\rightarrow s\overline{s} \; or \;  d\overline{d}) = 
\frac{ \alpha\tan^2\theta_H }{8 \cos^2\theta_W }   
\left( \frac{5}{9} 
\right)M_{Z'_t}.
\eeq

\beq
\Gamma_{\bf{II}}(Z'_t\rightarrow \mu\overline{\mu} ) = 
\frac{ \alpha\tan^2\theta_H }{8 \cos^2\theta_W }   
\left( \frac{5}{3} 
\right)M_{Z'_t}.
\eeq

\beq
\Gamma_{\bf{II}}(Z'_t\rightarrow \nu_\mu\overline{\nu_\mu}) = 
\frac{ \alpha\tan^2\theta_H }{8 \cos^2\theta_W }   
\left( \frac{1}{3} 
\right)M_{Z'_t}.
\eeq

\noindent
The partial widths for Model III are the
same as Model II, with the replacement
$\tan\theta_H \rightarrow \cot\theta_H$.

We thus have
the
{\bf Model II} total width:
\bea
\Gamma_{\bf{II}} & = &
\frac{ \alpha M_{Z'_t} \cot^2\theta_H }{8 \cos^2\theta_W }
\left[
 \sqrt{ 1 - \frac{4m_t^2}{M_{Z'_t}^2 }} 
\left( \frac{17}{9} - \frac{41 m_t^2}{9 M_{Z'_t}^2} 
\right) 
- \frac{17}{9} +
(2 +
\tan^4\theta_H )
\left( \frac{17}{9} + \frac{5}{9} + \frac{5}{3} +  \frac{1}{3}
\right)
\right]
\nonumber \\
& = & \frac{ \alpha M_{Z'_t} \cot^2\theta_H }{8 \cos^2\theta_W }
\left[
 \sqrt{ 1 - \frac{4m_t^2}{M_{Z'_t}^2 }} 
\left( \frac{17}{9} - \frac{41 m_t^2}{9 M_{Z'_t}^2} 
\right) 
+
\left( \frac{63}{9} \right)
+
\tan^4\theta_H 
\left( \frac{40}{9}
\right)
\right]
\label{eq_gamma_II}
\eea

\noindent
and the {\bf Model III} total width:
\bea
\Gamma_{\bf{III}} & = &
\frac{ \alpha M_{Z'_t} \cot^2\theta_H }{8 \cos^2\theta_W }
\left[
 \sqrt{ 1 - \frac{4m_t^2}{M_{Z'_t}^2 }} 
\left( \frac{17}{9} - \frac{41 m_t^2}{9 M_{Z'_t}^2} 
\right) 
+ 2\times\frac{17}{9} +
3\left( \frac{5}{9}_d + \frac{5}{3}_\ell +  \frac{1}{3}_{\nu\ell}\right)
\right]
\nonumber \\
& = & \frac{ \alpha M_{Z'_t} \cot^2\theta_H }{8 \cos^2\theta_W }
\left[
 \sqrt{ 1 - \frac{4m_t^2}{M_{Z'_t}^2 }} 
\left( \frac{17}{9} - \frac{41 m_t^2}{9 M_{Z'_t}^2} 
\right) 
+
\left( \frac{103}{9} \right)
\right]
\label{eq_gamma_III}
\eea

\noindent
Cross-sections are spin-color-summed on both
initial and final legs states.

\noindent
For {\bf{Model II}} the cross section is

\bea
\sigma_{\bf{II}}
& = &
\frac{9 \alpha^2 \pi  }{16\cos^4\theta_W}\cot^4\theta_H
\times \left(\frac{17}{9} \; \makebox{for initial
state $u+\bar{u}$;}, 
\; \frac{5}{9} \;\makebox{for initial $d+\bar{d}$ }\right)
\nonumber \\ 
& & \times
\left[ \beta(1 +\frac{1}{3}\beta^2)
 \times \left(\frac{17}{9} \; \makebox{for final $t+\bar{t}$ or $u+\bar{u}$;}, 
\; \frac{5}{9} \;\makebox{for final  $b+\bar{b}$ or $d+\bar{d}$ };
\right. \right. \nonumber \\
& & 
\qquad \qquad \left. \left.
\; \frac{5}{3} \;\makebox{for final $\tau+\bar{\tau}$ or $e+\bar{e}$ };
\; \frac{1}{3} \;\makebox{for final  $\nu_\tau+\bar{\nu_\tau}$ or
 $\nu_e+\bar{\nu_e}$ };
\right)
\right.
\nonumber \\ 
& & + \left. \frac{8}{9}\beta (1-\beta^2) \;  \makebox{(for
final  $t+\bar{t}$)}
\right] 
\left[ \frac{s}{(s-M_{Z'_t}^2)^2 + s\Gamma^2 } \right]\theta(s-4m_t^2)
\nonumber \\
\label{eq_sigma_II}
\eea

\noindent
For {\bf{Model III}} the cross section is

\bea
\sigma_{\bf{III}}
& = &
\frac{9 \alpha^2 \pi  }{16\cos^4\theta_W}\cot^4\theta_H
\times  \left(\frac{17}{9} \; \makebox{for initial $u+\bar{u}$;}, 
\; \frac{5}{9} \;\makebox{for initial $d+\bar{d}$ }\right)
\nonumber \\ 
& & \times
\left[ \beta(1 +\frac{1}{3}\beta^2)
\times \left(\frac{17}{9} \; \makebox{for final $t+\bar{t}$ or $u+\bar{u}$
or $c+\bar{c}$ ;}, 
\; \frac{5}{9} \;\makebox{for final $b+\bar{b}$ or $d+\bar{d}$ 
or $s+\bar{s}$ };
\right. \right. \nonumber \\
& & 
\qquad \qquad \left. \left.
\; \frac{5}{3} \;\makebox{for final $\tau+\bar{\tau}$ or $e+\bar{e}$
or $\mu+\bar{\mu}$ };
\; \frac{1}{3} \;\makebox{for final  $\nu_\tau+\bar{\nu_\tau}$ or
 $\nu_e+\bar{\nu_e}$ or   $\nu_\mu+\bar{\nu_\mu}$};
\right)
\right.
\nonumber \\ 
& & + \left. \frac{8}{9}\beta (1-\beta^2) \; \makebox{(for 
final $t+\bar{t}$)}
\right] 
\left[ \frac{s}{(s-M_{Z'_t}^2)^2 + s\Gamma^2 } \right]\theta(s-4m_t^2)
\nonumber \\
\label{eq_sigma_III}
\eea

\noindent
Dilepton final states are no doubt more sensitive 
discovery channels than quark dijets, or top for Models
I, II and III.

\vskip 0.5in
\noindent
{\bf \Large (B) Leptophobic Non-Standard Topcolor $Z'_t$ }
\vskip 0.05in
\noindent

Further 
non-standard models can be constructed 
for topcolor tilting with a
leptophobic interaction. Anomaly cancellation
is most easily implemented by having an
overall vector-like interaction, but with different
generations playing the role of anomaly
vector-like pairing. We do not mix
with the $U(1)_Y$ in these theories, but
we do normalize the coupling to the SM coupling $g_1$
as a convention.  

\noindent
({\bf{Model IV}}): 
quark generations $(1,3) \supset U(1)_2$

\noindent 
The dominant part of the interaction Lagrangian for Model 
IV is:
\bea
L'{}_{IV} & = & 
(\half g_1\cot\theta_H)Z_t'^{\mu}\left(  
\bar{t}_L\gamma_\mu t_L 
+\bar{b}_L\gamma_\mu b_L
+ f_1\bar{t}_R\gamma_\mu  t_R 
+ f_2\bar{b}_R\gamma_\mu  b_R 
\right. \nonumber \\
& &  \left.
-\bar{u}_L\gamma_\mu u_L 
-\bar{d}_L\gamma_\mu d_L
-f_1\bar{u}_R\gamma_\mu  u_R 
-f_2\bar{d}_R\gamma_\mu  d_R 
\right)
\eea
\noindent
Note that for topcolor tilting, we would require
the following: 
$f_1 > 0$ (attractive $\bar{t}t$ channel)
and/or $f_2 <0 $ (repulsive $\bar{b}b$ channel).
Also, $\cot\theta_h >> 1$ to avoid fine-tuning.

Hence, the cross-sections (spin-color-summed on both
initial and final legs states) for 
{\bf{Model IV}} are
\bea
\sigma
& = &
\frac{9 \alpha^2 \pi  }{16\cos^4\theta_W}\cot^4\theta_H
\times \left((1+f_1^2) \; \makebox{for initial
state $u+\bar{u}$;}, 
\; (1+f_2^2) \;\makebox{for initial $d+\bar{d}$ }\right)
\nonumber \\ 
& & \times
\left[ \beta(1 +\frac{1}{3}\beta^2)
 \times \left( (1+f_1^2)
 \; \makebox{for final $t+\bar{t}$ or $u+\bar{u}$;}, 
\; (1+f_2^2) \;\makebox{for final  $b+\bar{b}$ or $d+\bar{d}$ }\right)
\right.
\nonumber \\ 
& & + \left. f_1 \beta (1-\beta^2) \;  \makebox{(for
final  $t+\bar{t}$)}
\right] 
\left[ \frac{s}{(s-M_{Z'_t}^2)^2 + s\Gamma^2 } \right]\theta(s-4m_t^2)
\nonumber \\
\eea

\noindent
The partial widths for {\bf{Model IV}} are
\beq
\Gamma_{\bf{IV}}(Z'_t\rightarrow t\overline{t}) = 
\frac{ \alpha\cot^2\theta_H }{8 \cos^2\theta_W } \sqrt{ M_{Z'_t}^2 - 4m_t^2 } 
\left( (1+f_1^2)
\left[ 1-\frac{m_t^2}{M_{Z'_t}^2} 
\right] 
- 3f_1 
\left[ \frac{m_t^2}{M_{Z'_t}^2} \right]  \right)
\eeq

\beq
\Gamma_{\bf{IV}}(Z'_t\rightarrow u\overline{u}) = 
\frac{ \alpha\cot^2\theta_H M_{Z'_t}}{8 \cos^2\theta_W }  
\left( (1+f_1^2)
\right)
\eeq

\beq
\Gamma_{\bf{IV}}(Z'_t\rightarrow b\overline{b}) = 
\frac{ \alpha\cot^2\theta_H M_{Z'_t}}{8 \cos^2\theta_W }  
\left( (1+f_2^2)
\right)
\eeq

\beq
\Gamma_{\bf{IV}}(Z'_t\rightarrow d\overline{d}) = 
\frac{ \alpha\cot^2\theta_H M_{Z'_t}}{8 \cos^2\theta_W }  
\left( (1+f_2^2)
\right)
\eeq

\noindent
The total decay width for {\bf{Model IV}} is
\beq
\Gamma_{\bf{IV}} = 
\frac{ \alpha\cot^2\theta_H M_{Z'_t}}{8 \cos^2\theta_W } 
\left[ \sqrt{ 1 - \frac{4m_t^2}{M_{Z'_t}^2} } 
\left( (1+f_1^2) - (1+f_1^2+3f_1)\frac{m_t^2}{M_{Z'_t}^2} 
\right) 
+ (3+f_1^2 + 2f_2^2)
\right]
\eeq

\noindent
As a simple parameter scheme, leptophobic, b$_r$-phobic,
top$_r$-phyllic,  take $f_1=1$ and $f_2=0$:
\beq
\Gamma_{\bf{IV}}\rightarrow  
\frac{ \alpha\cot^2\theta_H M_{Z'_t}}{8 \cos^2\theta_W } 
\left[ \sqrt{ 1 - \frac{4m_t^2}{M_{Z'_t}^2} } 
\left( 2 - 5\frac{m_t^2}{M_{Z'_t}^2} 
\right) 
+ 4
\right]
\label{eq_gamma_IV}
\eeq

\bea
\sigma_{\bf{IV}}
& \rightarrow &
\frac{9 \alpha^2 \pi  }{16\cos^4\theta_W}\cot^4\theta_H
\times \left(2 \; \makebox{for initial
state $u+\bar{u}$;}, 
\; (1) \;\makebox{for initial $d+\bar{d}$ }\right)
\nonumber \\ 
& & \times
\left[ \beta(1 +\frac{1}{3}\beta^2)
 \times \left( 2
 \; \makebox{for final $t+\bar{t}$ or $u+\bar{u}$;}, 
\; (1) \;\makebox{for final  $b+\bar{b}$ or $d+\bar{d}$ }\right)
\right.
\nonumber \\ 
& & + \left. (1) \beta (1-\beta^2) \;  \makebox{(for
final  $t+\bar{t}$)}
\right] 
\left[ \frac{s}{(s-M_{Z'_t}^2)^2 + s\Gamma^2 } \right]\theta(s-4m_t^2)
\nonumber \\
\label{eq_sigma_IV}
\eea

\section{Cross Section at the Tevatron}

The total cross section for 
$p\bar{p} \rightarrow Z'_t \rightarrow t\bar{t}$ is
\begin{equation}
\sigma = \int_{0}^{\infty} \frac{d\sigma}{dm} dm
\label{eq_tot_xsec}
\end{equation}
where $d\sigma/dm$, the differential cross section at 
$t\bar{t}$ invariant mass $m$, is given by 
\begin{equation}
\frac{d\sigma}{dm} = \frac{2}{m} \int_{-\ln(\sqrt{s}/m)}^{\ln(\sqrt{s}/m)} dy_b \ \tau 
{\cal L}(x_p,x_{\bar{p}}) \ \hat{\sigma}(q\bar{q} \rightarrow Z'_t \rightarrow t\bar{t}). 
\label{eq_xsec}
\end{equation}
Here $\hat{\sigma}(q\bar{q} \rightarrow Z'_t \rightarrow t\bar{t})$ 
is the parton level subprocess cross section. The kinematic variable
$\tau$ is related to the initial state parton fractional momenta inside
the proton $x_p$ and anti-proton $x_{\bar{p}}$ by 
$\tau = x_p x_{\bar{p}} = m^2/s$.  The boost of the partonic system 
$y_b$ is given by $y_b = (1/2) \ln(x_p/x_{\bar{p}})$.
The partonic ``luminosity function'' is just the product of parton
distribution functions:
\begin{equation}
{\cal L}(x_p,x_{\bar{p}}) = q(x_p,\mu)\bar{q}(x_{\bar{p}},\mu) + \bar{q}(x_p,\mu)q(x_{\bar{p}},\mu)
\end{equation}
where $q(x, \mu)$ ($\bar{q}(x,\mu)$) is the parton distribution function of a 
quark (anti-quark) evaluated at fractional momenta $x$ and renormalization 
scale $\mu$.

The subprocess cross sections in 
equations~\ref{eq_sigma_I},~\ref{eq_sigma_II},~\ref{eq_sigma_III} 
and~\ref{eq_sigma_IV} are for spin and color
summing on both initial and final state legs, while most parton distributions
assume spin and color averaged on the initial state legs and spin and color
summing on the final state legs.  Therefore the subprocess cross sections 
given by equations~\ref{eq_sigma_I},~\ref{eq_sigma_II},~\ref{eq_sigma_III} 
and~\ref{eq_sigma_IV} must be multiplied by a factor of
\begin{equation}
\left(\frac{1}{spins}\right)^2\left(\frac{1}{colors}\right)^2 = 
\left(\frac{1}{2}\right)^2\left(\frac{1}{3}\right)^2 = \frac{1}{36}
\label{eq_spin_color}
\end{equation}
when used with parton distributions from PDFLIB~\cite{ref_pdflib} and other standard sources.
We have taken this into account when calculating the cross section.
we have also used $m_t=175$ GeV/c$^2$, and $\cos^2\theta_W=.768$.

\section{Width}

The minimum width of the $Z'_t$ depends on which model is
chosen.  For model I and II the minimum possible width imposed by 
equations~\ref{eq_gamma_I} and~\ref{eq_gamma_II} is around $\Gamma=0.016M$, 
with the actual 
minimum value depending on the $Z'_T$ mass. For Models III and IV there are no 
minimum widths imposed by equations~\ref{eq_gamma_III} and~\ref{eq_gamma_IV}
respectively. For model I the minimum possible width is of interest, 
because the cross section increases as the width decreases. Conversely, for 
models II, III and IV the cross section increases as the width increases, and 
the minimum possible width is of less interest.  All four models permit a 
width of $\Gamma=0.02M$. This width qualifies as a narrow resonance, since it is 
significantly less than the CDF detector resolution for $t\bar{t}$.  We will
also see that this width gives a significant cross section at the Tevatron 
for model IV, making it experimentally accessible. Therefore, we will 
concentrate on a width of $\Gamma=0.02M$ for the purpose of comparing cross 
sections among models and tabulating results. Table~\ref{tab_width} shows
how this width relates to the fundamental coupling parameter $\cot^2\theta_H$.
\begin{table}[tbh]
\begin{center}
\begin{tabular}{|c||c|c|c|c|}\hline 
Mass & 
\multicolumn{4}{c|}{$\cot^2\theta_H$ for Model} \\ \cline{2-5}
(GeV/$c^2$) & I & II & III & IV \\ \hline
400      & 4.56 & 1.78 & 1.33 & 3.52 \\
500      & 3.87 & 1.66 & 1.28 & 3.18 \\
600      & 3.53 & 1.59 & 1.25 & 3.00 \\
700      & 3.34 & 1.55 & 1.24 & 2.90 \\
800      & 3.23 & 1.52 & 1.22 & 2.84 \\
$\infty$ & 2.87 & 1.44 & 1.19 & 2.64 \\
\hline
\end{tabular}
\end{center}
\caption[Mixing angle $\cot^2\theta_H$]{ 
As a function of $Z'_t$ mass, we tabulate  
the value of $\cot^2\theta_H$ for a width of $\Gamma=0.02M$ for
models I - IV.}
\label{tab_width}
\end{table}

\section{Numerical Results for the Tevatron}

\begin{figure}[tbh]
\centerline{\epsfig{figure=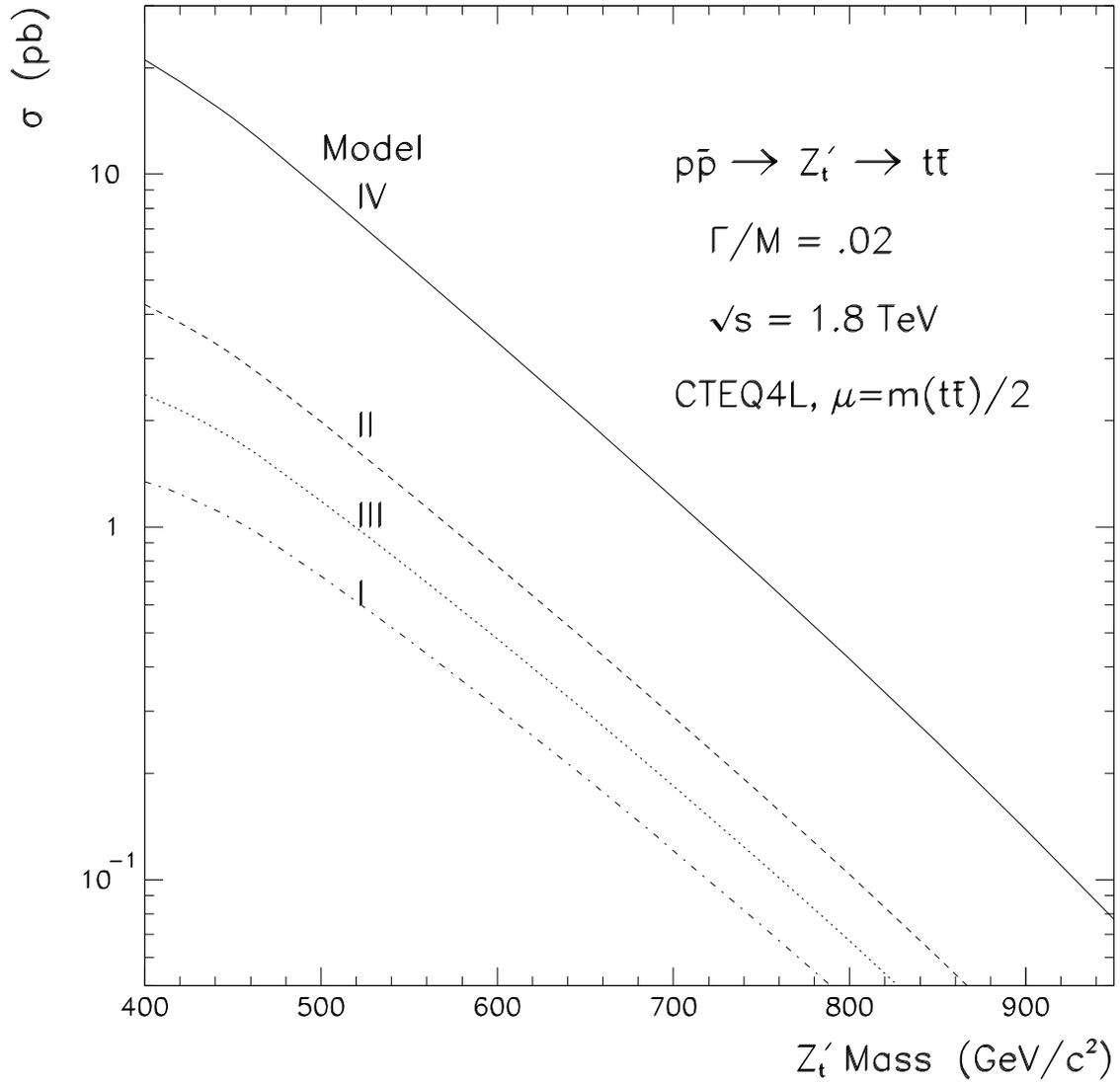,width=6.5in}}
\caption{ 
The lowest order cross section for the process 
$p\bar{p} \rightarrow Z'_t \rightarrow t\bar{t}$ from table~\ref{tab_xsec}.}
\label{fig_xsec}
\end{figure}

We have calculated the lowest order cross section for the process
$p\bar{p} \rightarrow Z'_t \rightarrow t\bar{t}$ using a 
computer program that numerically performs the integrations in 
equations \ref{eq_tot_xsec} and 
\ref{eq_xsec}. The integration in Eq.~\ref{eq_tot_xsec} was performed using 
the mass interval $M-10\Gamma<m<M+10\Gamma$.
For Models I, II, III and IV we used subprocess cross sections 
~\ref{eq_sigma_I},~\ref{eq_sigma_II},~\ref{eq_sigma_III} 
and~\ref{eq_sigma_IV} multiplied by the spin-color factor in 
equation~\ref{eq_spin_color}.
The only parameter of the topcolor model that affects the 
cross section is the mixing angle 
$\cot^2\theta_H$, or equivalently the width $\Gamma$ which is 
related to it. After the width choice has been made, the 
only uncertain parameters of the calculation are the choice of parton 
distributions and renormalization scale $\mu$. For a default parton 
distribution set we have chosen CTEQ4L~\cite{ref_cteq}. This is a modern parton 
distribution set appropriate for leading order calculations and is available 
in PDFLIB~\cite{ref_pdflib}. For a default renormalization scale we choose 
$\mu=m/2$, half the $t\bar{t}$  invariant mass.  This scale has the 
benefit that it reduces to the usual $\mu=m_t$ at top production threshold, 
but also increases with increasing $t\bar{t}$ invariant mass. With these
choices, the total cross section for 
$p\bar{p} \rightarrow Z'_t \rightarrow t\bar{t}$ for a $Z'_t$
width of $\Gamma=0.02M$ is tabulated in table~\ref{tab_xsec} and displayed in 
Fig.~\ref{fig_xsec} for each of Models I through IV. 

\begin{figure}[tbh]
\centerline{\epsfig{figure=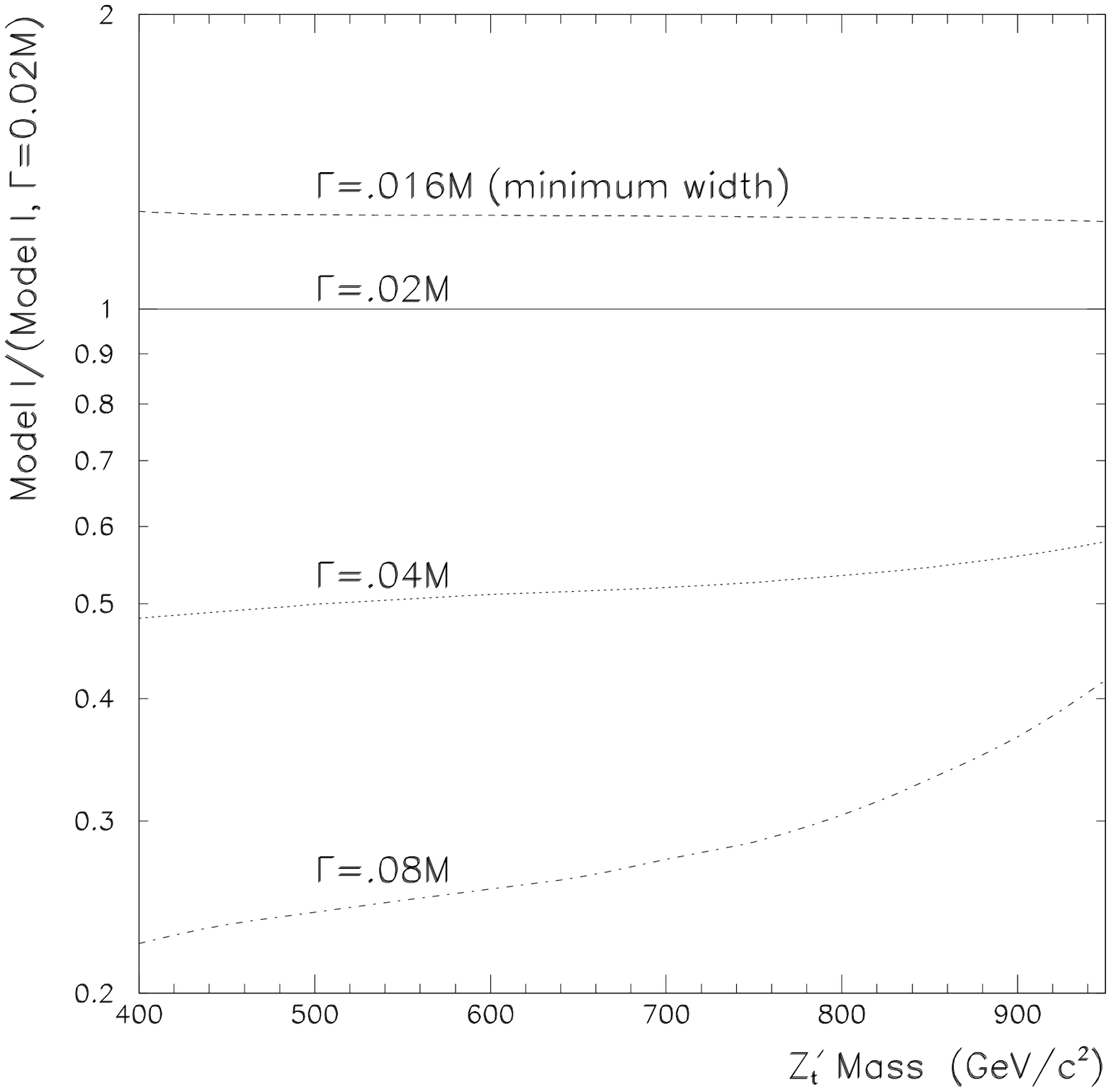,width=6.5in}}
\caption{ 
For Model I we plot the $Z'_t$ cross section for various widths 
divided by the cross section for a width $\Gamma=.02M$.}
\label{fig_xsec_model1}
\end{figure}

\begin{figure}[tbh]
\centerline{\epsfig{figure=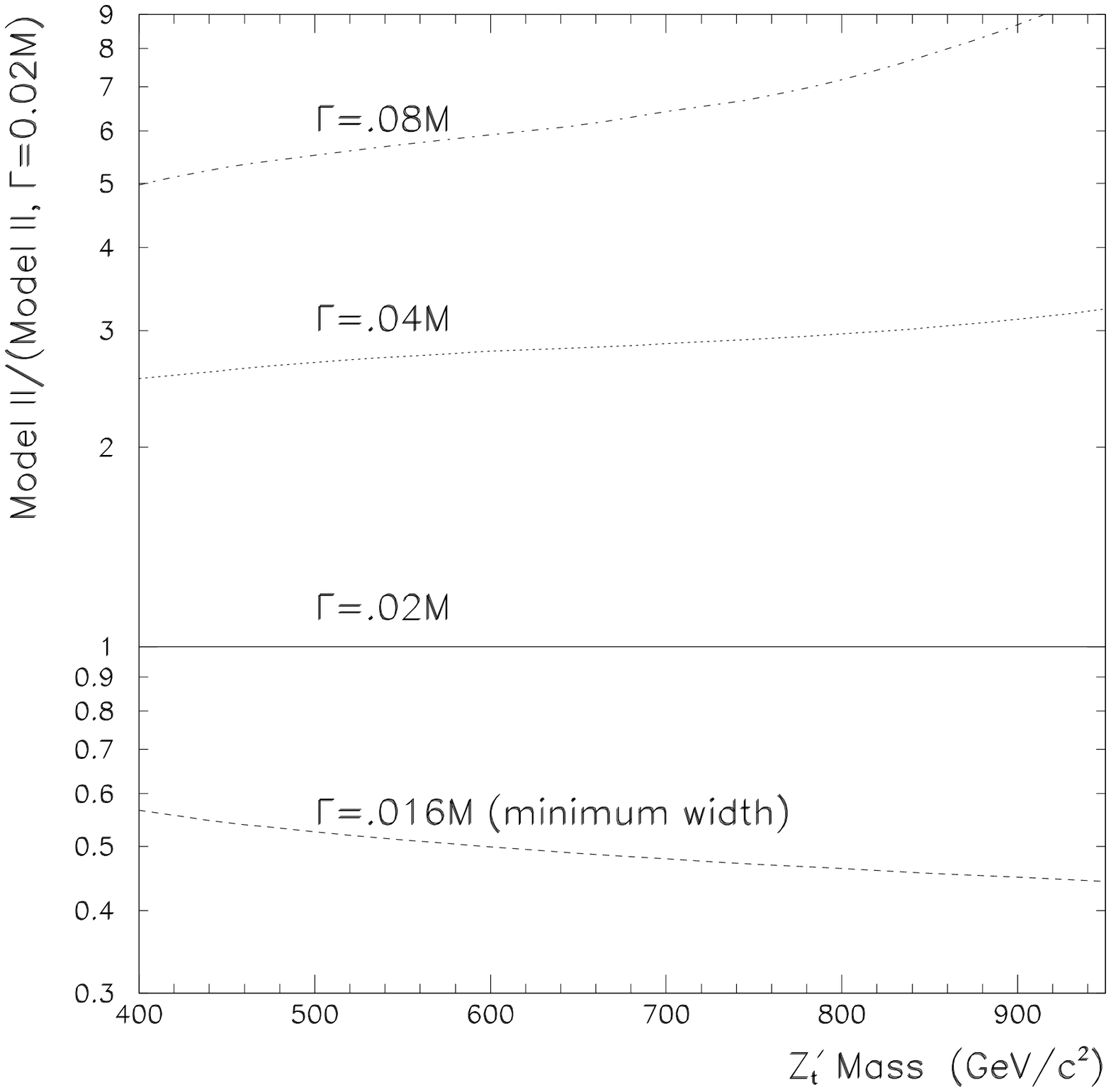,width=6.5in}}
\caption{ 
For Model II we plot the $Z'_t$ cross section for various widths 
divided by the cross section for a width $\Gamma=.02M$.}
\label{fig_xsec_model2}
\end{figure}

\begin{figure}[tbh]
\centerline{\epsfig{figure=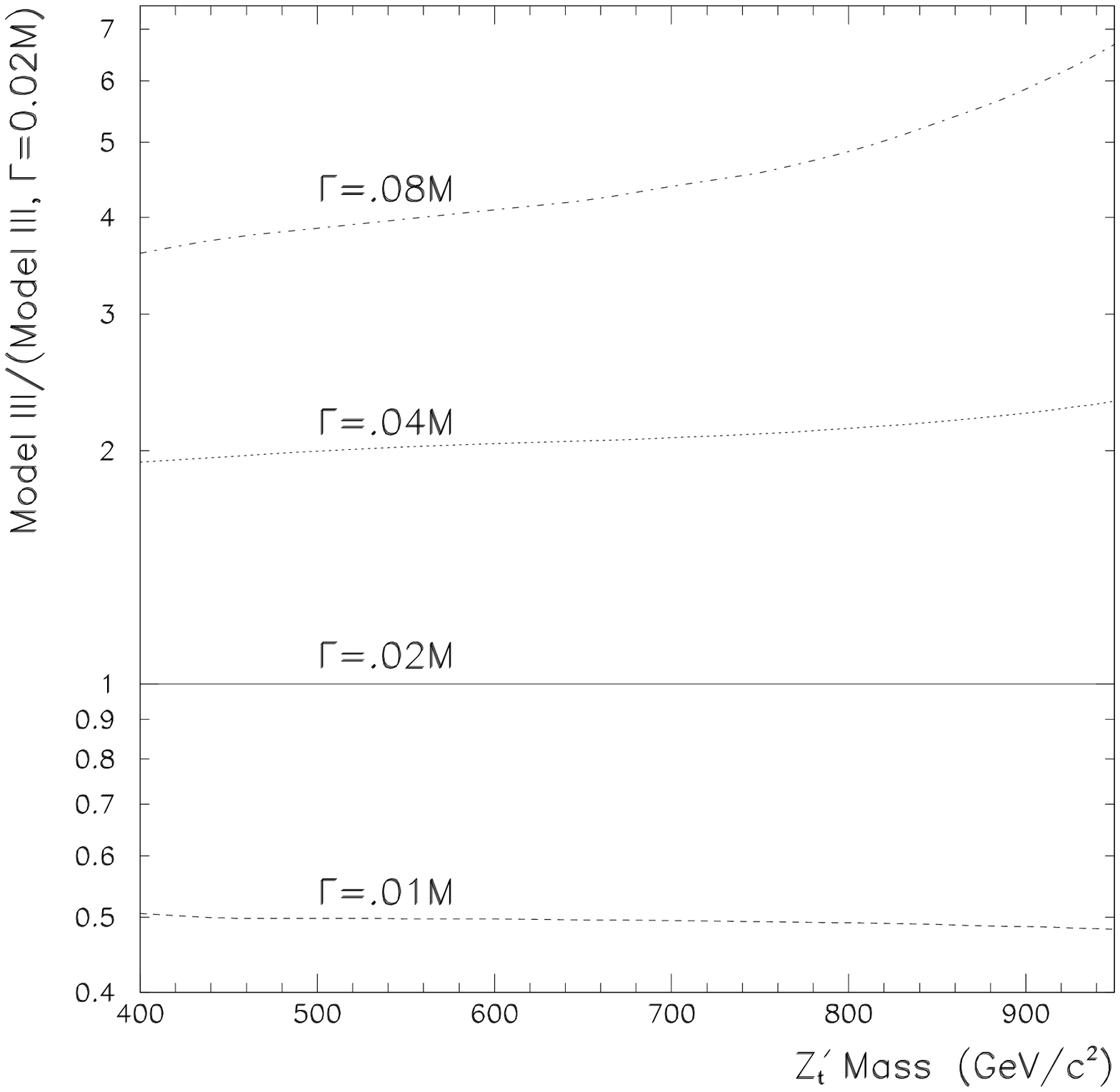,width=6.5in}}
\caption{ 
For Model III we plot the $Z'_t$ cross section for various widths 
divided by the cross section for a width $\Gamma=.02M$.}
\label{fig_xsec_model3}
\end{figure}

\begin{figure}[tbh]
\centerline{\epsfig{figure=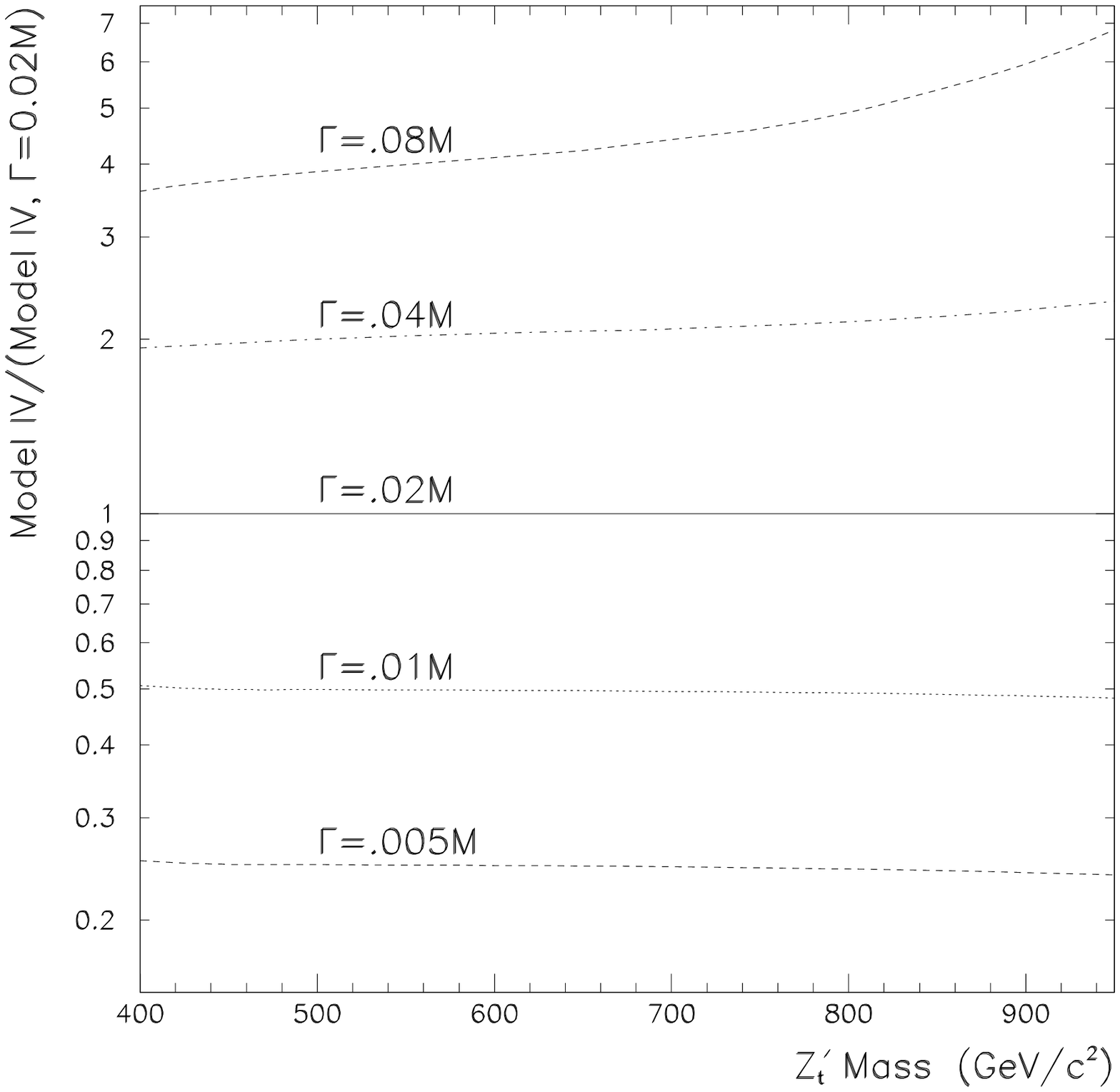,width=6.5in}}
\caption{ 
For Model IV we plot the $Z'_t$ cross section for various widths 
divided by the cross section for a width $\Gamma=.02M$.}
\label{fig_xsec_model4}
\end{figure}

\begin{figure}[tbh]
\centerline{\epsfig{figure=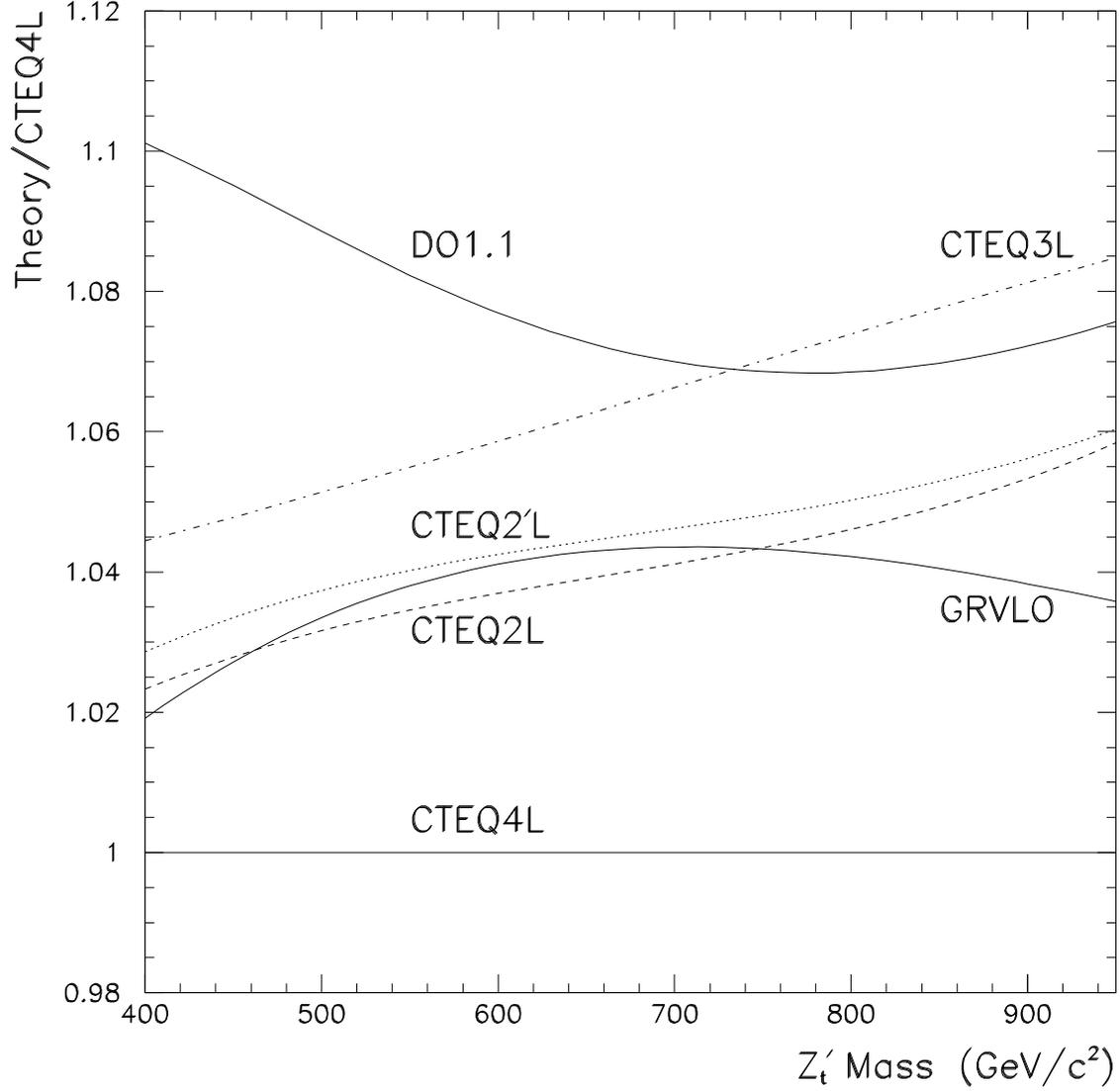,width=6.5in}}
\caption{ 
The $Z'_t$ cross section for Model II with width $\Gamma=0.02M$ for
various choices of parton distribution
function divided by the cross section with CTEQ4L parton distribution
functions. The different choices are
CTEQ2L, CTEQ2$'$L, CTEQ3L~\cite{ref_cteq}, 
DO1.1~\cite{ref_do} and GRVLO~\cite{ref_grv}.}
\label{fig_xsec_pdf}
\end{figure}

\begin{figure}[tbh]
\centerline{\epsfig{figure=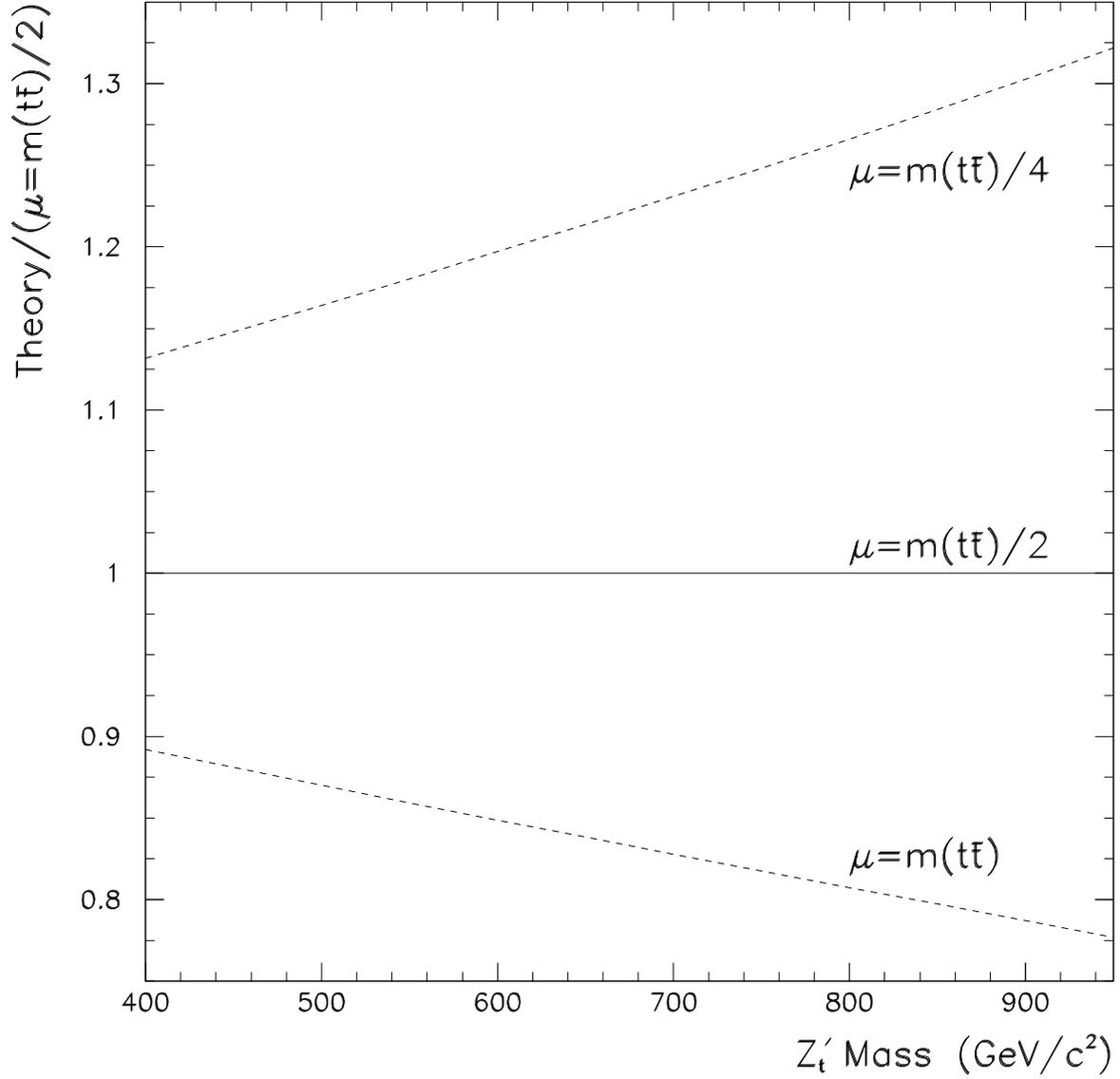,width=6.75in}}
\caption{ 
The $Z'_t$ cross section for Model II with width $\Gamma=0.02M$ with 
two other choices of renormalization scale 
divided by the cross section using renormalization scale $\mu=m/2$.
The two choices are $\mu=m$ and $\mu=m/4$.}
\label{fig_xsec_mu}
\end{figure}

We have explored the variation in cross section when changing the 
$Z'_t$ model, the $Z'_t$ width, and when changing the parton 
distributions and renormalization scale. 
Figure~\ref{fig_xsec_model1} shows that for Model I 
the cross section is approximately inversely proportional to the width.
Figures ~\ref{fig_xsec_model2}, ~\ref{fig_xsec_model3}, 
~\ref{fig_xsec_model4} shows that for Models II through IV the cross section 
is approximately proportional to the width.
The variation in cross section when changing the parton distribution 
functions is displayed in Fig~\ref{fig_xsec_pdf}. We have only included
parton distributions determined in the 1990's and extracted at lowest order, 
appropriate for our lowest order calculation. By coincidence, the choice of 
CTEQ4L happens to yield a lower cross section than the others and is therefore
also a conservative choice.  
The variation in cross section when increasing or decreasing the 
renormalization scale is shown in Fig~\ref{fig_xsec_mu}. 

The cross section for
the $Z'_t$ in Model IV is large enough that it should be possible to observe
or exclude this model, for a significant range of masses and widths, using 
current data from the Tevatron Collider. Preliminary results on a search for
narrow resonances decaying to $t\bar{t}$ are available from CDF
~\cite{ref_cdf} and can be used to constrain a $Z'_t$ from Model IV.
We apologize for an error in the predicted cross section for the standard 
$Z'_t$ in the preliminary CDF search. The predictions for the $Z'_t$ presented 
here supersedes those presented in reference~\cite{ref_cdf}. We eagerly 
anticipate the next run of the Tevatron Collider, which should be sensitive 
to the $Z'_t$ in all the models we have proposed.

\begin{table}[tbh]
\begin{center}
\begin{tabular}{|c||l|l|l|l|}\hline 
Mass & \multicolumn{4}{c|}
{$\sigma(p\bar{p} \rightarrow Z'_t \rightarrow t\bar{t})$ [pb]}\\ 
\cline{2-5}
(GeV/$c^2$) & Model I & 
              Model II & 
              Model III & 
              Model IV \\ \hline 
400 & $1.34$ &
      $4.27$ & 
      $2.37$ &
      $2.10\times 10^1$ \\
450 & $1.05$ &
      $3.07$ &
      $1.78$ &
      $1.44\times 10^1$ \\
500 & $7.23\times 10^{-1}$ &
      $1.98$ &
      $1.18$ &
      $8.97$ \\
550 & $4.77\times 10^{-1}$ &
      $1.25$ &
      $7.61\times 10^{-1}$ & 
      $5.48$ \\
600 & $3.07\times 10^{-1}$ &
      $7.73\times 10^{-1}$ &
      $4.81\times 10^{-1}$ &
      $3.33$ \\
650 & $1.94\times 10^{-1}$ &
      $4.76\times 10^{-1}$ &
      $3.00\times 10^{-1}$ &
      $2.01$ \\
700 & $1.21\times 10^{-1}$ &
      $2.89\times 10^{-1}$ &
      $1.85\times 10^{-1}$ &
      $1.21$ \\
750 & $7.42\times 10^{-2}$ &
      $1.74\times 10^{-1}$ &
      $1.12\times 10^{-1}$ &
      $7.18\times 10^{-1}$ \\
800 & $4.47\times 10^{-2}$ &
      $1.03\times 10^{-1}$ &
      $6.70\times 10^{-2}$ &
      $4.22\times 10^{-1}$ \\
850 & $2.64\times 10^{-2}$ &
      $6.03\times 10^{-2}$ &
      $3.94\times 10^{-2}$ &
      $2.44\times 10^{-1}$ \\
900 & $1.53\times 10^{-2}$ &
      $3.45\times 10^{-2}$ &
      $2.27\times 10^{-2}$ &
      $1.39\times 10^{-1}$ \\
950 & $8.64\times 10^{-3}$ &
      $1.93\times 10^{-2}$ &
      $1.27\times 10^{-2}$ &
      $7.72\times 10^{-2}$ \\
\hline
\end{tabular}
\end{center}
\caption[Cross section]{ 
As a function of $Z'_t$ mass in Models I-IV, we tabulate the cross section for the
process $p\bar{p} \rightarrow Z'_t \rightarrow t\bar{t}$ at 
$\sqrt{s}=1.8$ TeV for a $Z'_t$ width of $\Gamma=0.02M$ using 
CTEQ4L parton distributions and a renormalization scale $\mu=m/2$.
(half the invariant mass of the $t\bar{t}$ system.)}
\label{tab_xsec}
\end{table}

\clearpage

\clearpage

\clearpage

\clearpage

\clearpage

\clearpage

\clearpage

\end{document}